\documentclass[9pt,twocolumn,twoside]{osajnl}
\pdfoutput=1

\journal{josab}

\setboolean{shortarticle}{false}

\title{Starting Dynamics of a Fully Electronically Controlled Linear Mamyshev Oscillator}

\author[1,*]{Yi-Hao Chen}
\author[1]{Pavel Sidorenko}
\author[1]{Frank Wise}

\affil[1]{School of Applied and Engineering Physics, Cornell University, Ithaca NY 14853, USA}
\affil[*]{Corresponding author: yc2368 @cornell.edu}

\usepackage{siunitx}
\newcommand{\SIadj}[2]{\SI[number-unit-product={\text{-}}]{#1}{#2}}
\newcommand{\SIrangeadj}[3]{\numrange[range-phrase=--]{#1}{#2}~\si{#3}}

\begin{abstract}
We start an environmentally-stable linear Mamyshev oscillator with electronically-controlled modulated pump and a moving filter. It delivers a \SIadj{21}{\nano\joule} pulse that becomes \SI{65}{\fs} in duration after a compressor. Reliable starting into stable mode-locking is found achievable with a modulated mode-locked state observed only when the modulation frequency is larger than \SI{70}{\kilo\Hz}. To avoid the damage of the gain fiber, two Faraday rotators are introduced. Besides, we have studied several aspects toward obtaining successful starting.
\end{abstract}

\setboolean{displaycopyright}{true}
\dates{}
\doi{}

\begin{document}

\maketitle

\section{Introduction}
Ultrafast fiber lasers have widespread use in several fields, such as machining and imaging \cite{Liu1997a,Xu2013,Hasegawa2019}. They have become an attractive alternative to solid-state lasers due to their waveguide nature, excellent thermal-optic properties, and robust operation. Recently, their performance become comparable to their solid-state counterparts \cite{Fu2018}.

In the past few years, a new type of mode-locked fiber lasers, which boosts the peak power to the megawatt level, has been developed. It realizes an effective saturable absorber (SA) with offset spectral filtering. A pulse undergoes nonlinear spectral broadening before it encounters a bandpass filter. A weak pulse that generates insufficient spectral broadening will be blocked by the bandpass filter with an offset wavelength while more-intense pulses are preferentially transmitted/reflected.

This idea of offset spectral filtering was proposed by P.~Mamyshev for pulse regeneration \cite{Mamyshev1998}. But it was only until recently that offset spectral filtering was applied for short-pulse generation \cite{Pitois2008}. Offset spectral filtering benefits from the sharp contrast of filter transmission/reflection and the tunability of filter wavelengths. In particular, the increase of filter separation shows a tendency for broadband pulses, which is consistent with the accumulation of large nonlinear phase and results in energetic short pulses. Pulses up to \SI{1}{\micro\joule} and \SIadj{41}{\fs} full-width at half-maximum (FWHM) duration with \SIadj{13}{\MW} peak power \cite{Liu2019}, as well as pulses with \SIadj{\sim400}{\nm} bandwidth and \SIadj{17}{\fs} FWHM duration \cite{Ma2020}, have been experimentally achieved. These so-called Mamyshev oscillators, whose SA is based on offset spectral filtering, can be built with polarization-maintaining (PM) fibers and yields environmental stability.

However, the Mamyshev SA has a disadvantage: the suppression of continuous-wave (CW) lasing due to offset filters naturally hinders the growth of fluctuations. As a result, starting a mode-locked state from noise becomes a major challenge. Overlapped filter passbands (where CW can coexist with the mode-locked state) allow to self-start a linear Mamyshev oscillator; however, this approach yields only relatively low-energy pulses due to a reduced filter separation \cite{Zeludevicius2018}. Nonlinear polarization evolution (NPE) can be coupled with the offset spectral filtering for self-starting, but this sacrifices environmental stability \cite{Ma2020}. In addition to allowing the coexistence of a CW state to ignite pulsation, several studies have instead used an external seed pulse or an acousto-optic modulator \cite{Liu2019,Regelskis2015,Liu2017}. Also, self-starting a ring Mamyshev oscillator with overlapped filter passbands has been achieved by modulating the pump \cite{Samartsev2017}. To avoid sacrificing pulse energy for starting with a reduced filter separation, a non-PM Q-switching arm was introduced to start a laser that generates \SIadj{190}{\nano\joule} pulses at \SI{1}{\micro\meter} \cite{Sidorenko2018}; the same technique was applied to obtain \SIadj{31}{\nano\joule} pulses at \SI{1550}{\nm} \cite{Olivier2019}. Although these techniques succeed in starting Mamyshev oscillators, they all compromise performance and/or environmental stability, or require a coherent seed pulse.

In general, a linear Mamyshev oscillator offers simplicity by requiring only less components of its ring counterpart. However, to date, linear Mamyshev oscillators have not come close to the performance achieved by ring-cavity designs. This then raises questions about (1) the limit of performance of the linear-cavity ones and, most importantly, (2) if it can be started without introducing additional drawbacks as mentioned above.

In this paper, we begin by investigating the linear Mamyshev oscillator \cite{Zeludevicius2018,Regelskis2015}. Although it shows good performance, we find it susceptible to damage of the gain fiber during starting. An improved design is then proposed to avoid the damage. With this design, we can reliably start a linear Mamyshev oscillator with modulated pump and a moving filter. The delivered pulse reaches more than 15 times peak power than prior linear Mamyshev oscillators. In addition, we study the starting dynamics with Dispersive Fourier Transform (DFT) \cite{Goda2013}.

\section{Preliminary Approach}

\begin{figure}[h!]
\centering\includegraphics[width=3.32in]{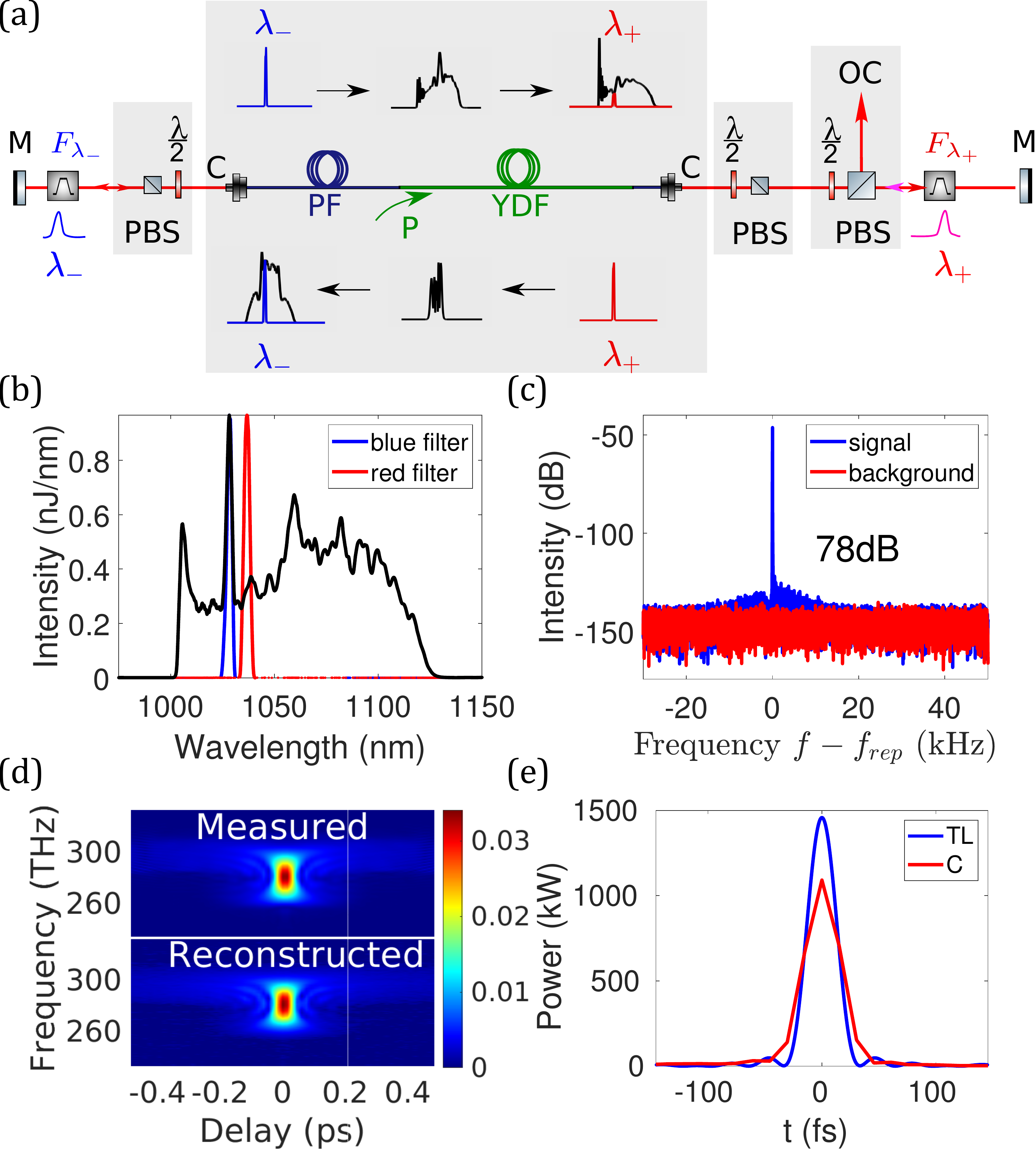}
\caption{(a) Schematic of linear Mamyshev oscillator. The spectra illustrate the Mamyshev saturable-absorber mechanism. The pump power is maintained at \SI{3}{\watt}. PF: passive fiber; YDF: Yb-doped fiber; $F_{\lambda_-}$: blue filter ($F_{\lambda_-}=\SI{1027}{\nm}$); $F_{\lambda_+}$: red filter; $\lambda/2$: half-wave plate; PBS: polarizing beam splitter; OC: output coupler (\SI{\sim85}{\%}); P: pump light from a \SIadj{976}{\nm} diode laser; C: collimator; M: mirror. (b) Spectra of \SIadj{47}{\nano\joule} pulses and the corresponding filters. (c) RF spectrum; $f_{\text{rep}}=\SI{16.8}{\mega\Hz}$: repetition rate of the laser (Video bandwidth, or VBS, $=\SI{5}{\Hz}$, resolution bandwidth, or RBS, $=\SI{10}{\Hz}$). (d) Measured and reconstructed FROG trace of the compressed compressed pulse. (e) Temporal profiles of the retrieved and the transform-limited pulse. TL: transform-limited pulse; C: compressed pulse.}
\label{fig:result_with_SBS}
\end{figure}

We started with the experimental setup of the linear Mamyshev oscillator in Fig.~\ref{fig:result_with_SBS}(a). The fibers have \SIadj{6}{\micro\meter} core diameters and are PM for environmental stability. Two interference bandpass filters with \SIadj{4}{\nm} FWHM bandwidth are employed for offset spectral filtering, which is illustrated with the spectral evolution in Fig.~\ref{fig:result_with_SBS}(a). The center wavelength of the blue one is fixed at \SI{1027}{\nm}, while that of the red one can be varied with a motorized rotation stage to attain \SIrangeadj{0}{15}{\nm} separation of filter center wavelengths.

To start this laser, two steps are applied: (1) the filter passbands are overlapped for self-starting and (2) the filter separation is subsequently increased for evolution to a high-energy single-pulse state. With a filter separation of \SI{6}{\nm}, the laser first self-starts in a multi-pulse state. By increasing the filter separation to \SI{12}{\nm}, we achieve a \SIadj{47}{nJ} single-pulse state with \SIadj{130}{\nm} bandwidth [Fig.~\ref{fig:result_with_SBS}(b)]. The pulse is compressed by a \SIadj{1000}{line\per\mm} grating pair to \SIadj{35}{\fs} FWHM duration and characterized with a frequency-resolved optical gating (FROG) \cite{Trebino2012} [Fig.~\ref{fig:result_with_SBS}(d, e)]. The corresponding peak power exceeds \SI{1}{\MW}. The radio-frequency (RF) spectrum exhibits \SIadj{78}{\dB} contrast [Fig.~\ref{fig:result_with_SBS}(c)], which represents stable mode-locking.

This linear cavity achieves the goal of offering excellent performance without sacrificing environmental stability or requiring an external coherent seed pulse. However, we sometimes observe catastrophic damage of the Yb-doped fibers while starting. We suspect that SBS is the reason for the damage. SBS has been observed in linear oscillators and is responsible for self-Q-switching or random pulsing \cite{VChernikov1997,Hideur2000,Upadhyaya2010a,Mallek2013,Hanzard2017}. Because we employ overlapped filter passbands for self-starting, SBS can occur. It can be uncontrollably amplified above the damage threshold in the gain fiber, which we suspect can cause irreversible damage. Further attempts to avoid damage by starting the laser with reduced pump power also failed.

It’s worth mentioning that recently Boulanger et al. have demonstrated a self-polarizing linear Mamyshev oscillator that delivers \SIadj{21}{\nano\joule} pulses at \SI{1550}{\nm} \cite{Boulanger2020}. They have started it with a reduced filter separation and a saturable absorber mirror. However, they didn’t claim to observe any damage while starting. More study on the damage mechanism with overlapped passbands is thus required.

\section{SBS-FREE LINEAR MAMYSHEV OSCILLATOR}

\subsection{Design and steps of starting}

\begin{figure}[h!]
\centering\includegraphics[width=3.32in]{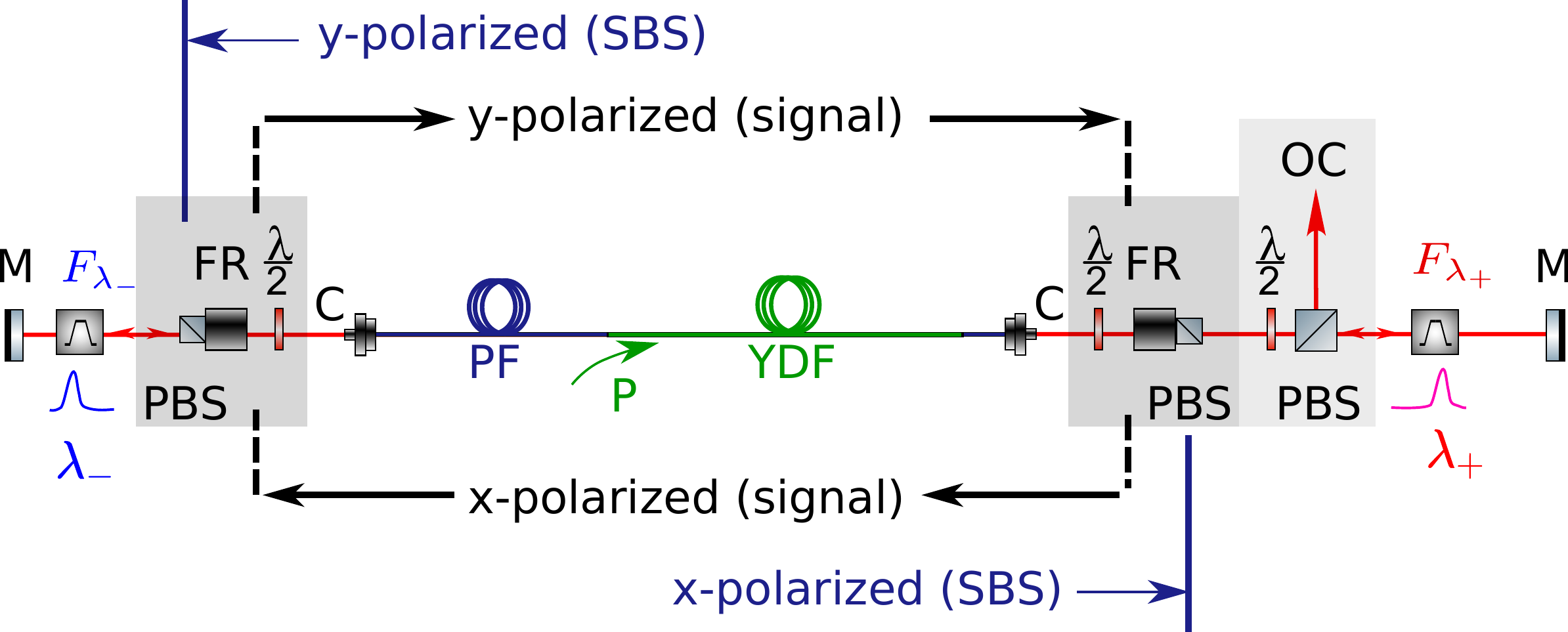}
\caption{Schematic of an SBS-free linear Mamyshev oscillator with FRs. The left passive fiber, including the pigtail of the collimator, is \SI{2}{\meter} while the right one is \SI{0.5}{\meter}; the YDF is \SI{3}{\meter}. FR: Faraday rotator.}
\label{fig:PM_linear_Mamyshev_linear_noSBS}
\end{figure}

To avoid damage from SBS, we built a linear oscillator whose pulses propagate with orthogonal polarizations in different directions (Fig.~\ref{fig:PM_linear_Mamyshev_linear_noSBS}). In figure \ref{fig:PM_linear_Mamyshev_linear_noSBS}, two Faraday rotators (FRs) are introduced to allow only y-polarized light propagating toward the output but x-polarized light in the opposite direction. SBS is mainly generated in a polarization that is linearly co-polarized with the pulse \cite{Agrawal2013}; as a result, it will be blocked by the polarizing beam splitters after the FRs.

Unfortunately, we fail to observe self-starting of the laser with the FRs, regardless of the filter separation. Although SBS is catastrophic for high-energy pulse generation, we conclude that it is essential for self-starting a PM linear Mamyshev oscillator. It acts as an intrinsic coherent-seed generator for self-seeding an oscillator. We observe successful self-starting only without FRs but not with them in the cavity.

\begin{figure}[h!]
\centering\includegraphics[width=2.28in]{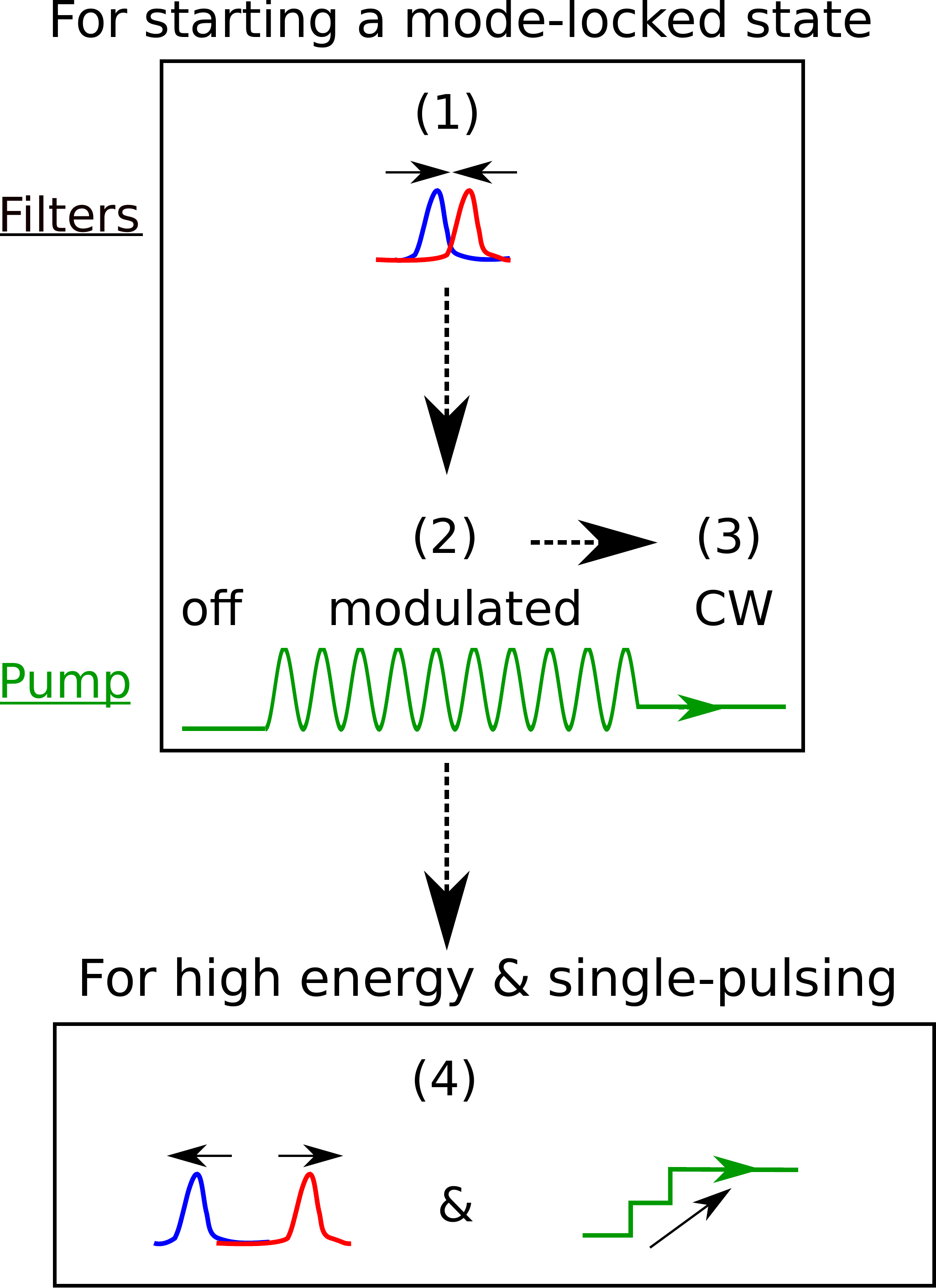}
\caption{The starting procedure of an SBS-free linear Mamyshev oscillator with pump modulation and a moving filter.}
\label{fig:procedure}
\end{figure}

Since self-starting can’t be observed with our laser, we adopt a strategy based on pump modulation for starting (Fig.~\ref{fig:procedure}) \cite{Samartsev2017}. (1) Initially, the laser requires overlapped filter passbands. (2, 3) We turn on and off the pump modulation to obtain a (possibly multi-pulsing) stable mode-locked state. (4) It later reaches high-energy single-pulsing states by increasing the filter separation and/or the pump power.

\subsection{Starting a mode-locked state}

\subsubsection{Pump modulation}

\begin{figure}[h!]
\centering\includegraphics[width=3.32in]{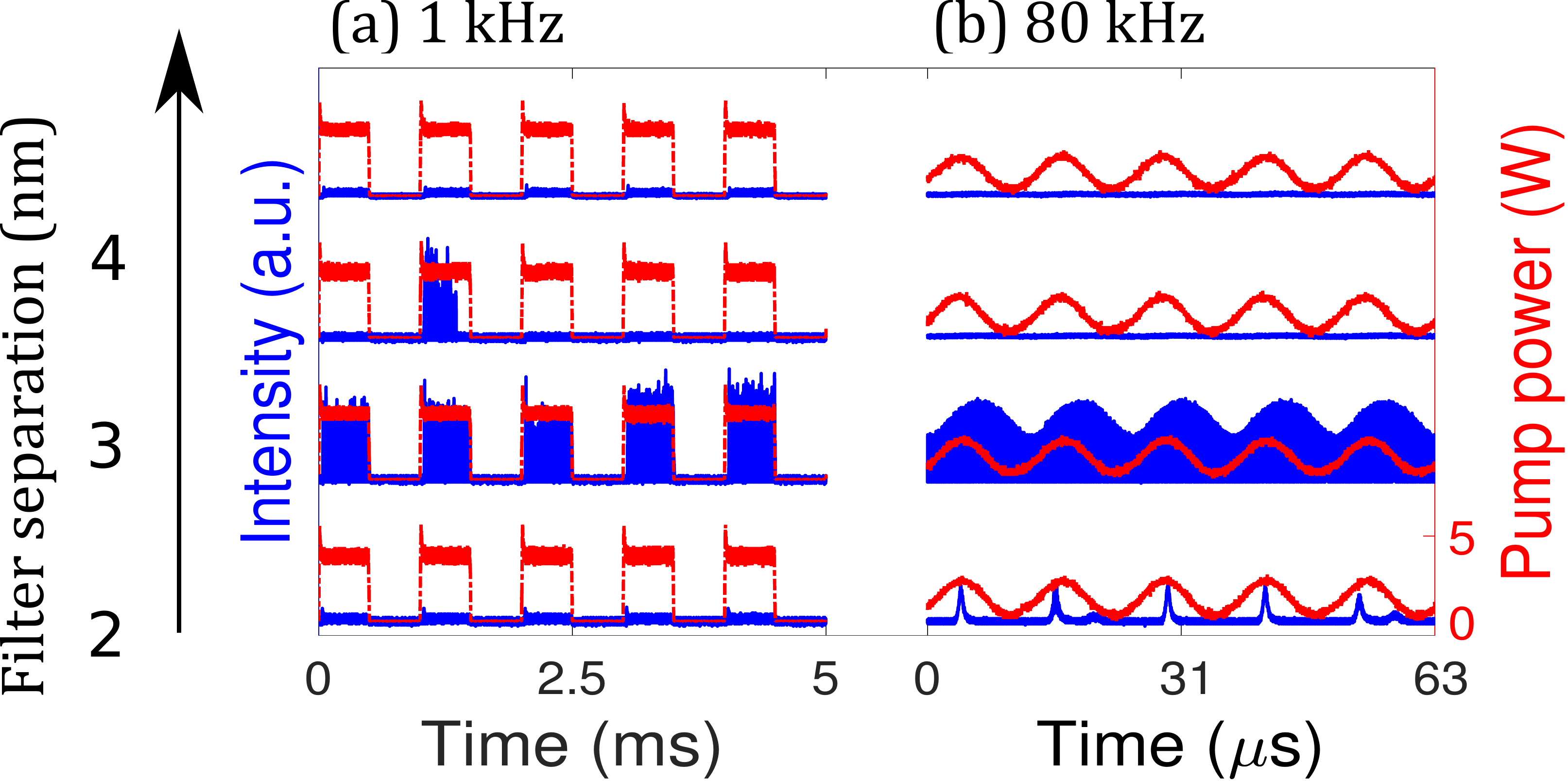}
\caption{Analysis of pump modulation with different filter separations. (a) The self-Q-switching state under \SIadj{1}{\kHz} modulation frequency and (b) the modulated mode-locked state under \SIadj{80}{\kHz} one.}
\label{fig:pump_modulation}
\end{figure}

We first studied the effect of pump modulation frequency on starting. For this work, we chose \SI{50}{\%} duty cycle and \SI{100}{\%} modulation depth for the pump. While several modulation frequencies have been studied, here we show only representative results for \SI{1}{\kHz} [Fig.~\ref{fig:pump_modulation}(a)] and \SI{80}{\kHz} [Fig.~\ref{fig:pump_modulation}(b)]. When the modulation frequency is below \SI{70}{\kHz}, we observe sporadic broadband self-Q-switching. This is more likely to be observed with larger pump power. Its fluctuating intensity results in fluctuating spectral bandwidth. When the modulation frequency is above \SI{70}{\kHz}, mode-locking occurs. Pulses are emitted at the cavity repetition rate, with the envelope of the pulse train modulated by the pump. In contrast to self-Q-switching, this state exhibits a consistent behavior throughout pump modulation where the pulses continues propagating in the cavity without dying out.

We also studied the influence of filter separations on the laser behaviors. Fig.~\ref{fig:pump_modulation} shows that we can obtain self-Q-switching and modulated mode-locked state only for filter separations near \SI{3}{\nm}. With larger separations, CW breakthrough is hindered, and only amplified stimulated emission remains. For smaller separations, CW lasing occurs because it is the condition of the least loss in the cavity. Besides appropriate filter separations, large modulation depth is also found helpful to initiate pulsation.

Based on these results, we can reliably achieve mode-locking with \SIadj{80}{\kHz} modulation frequency and \SIadj{3.4}{\nm} filter separation. In contrast to frequent failure of obtaining a stable mode-locked state with self-Q-switching at low modulation frequency, this modulated mode-locked state is found to evolve reliably into a stable mode-locked state as modulation is turned off to create CW pump.

\subsubsection{Various States During Starting}

\begin{figure}[h!]
\centering\includegraphics[width=3.32in]{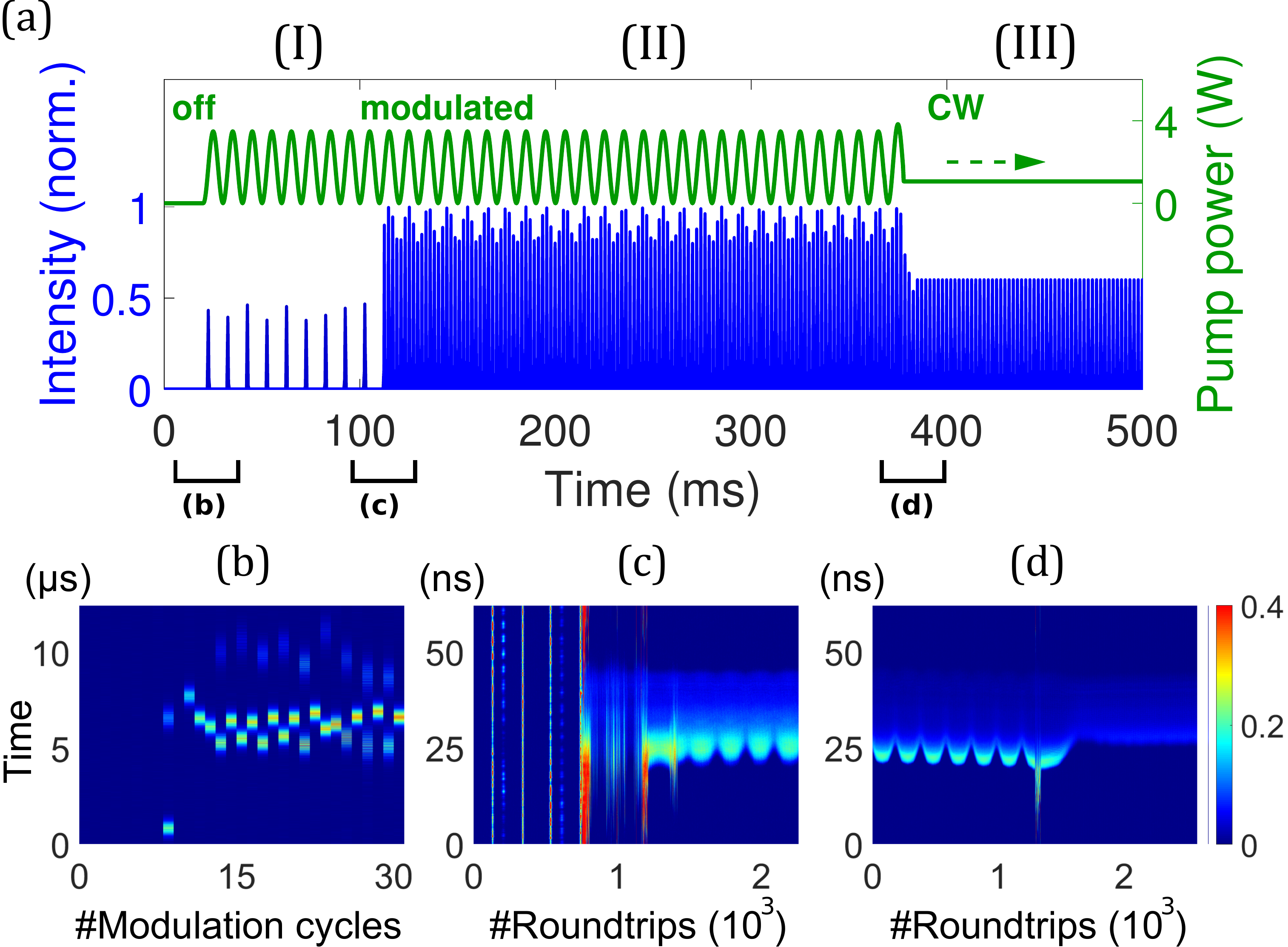}
\caption{Starting dynamics with pump modulation. Panel (a) illustrates the concept and the scale is not quantitative. (I) Temporary self-Q-switched state appears after turning on the \SIadj{80}{\kHz} modulated pump and then (II) it spontaneously evolves into a periodically modulated mode-locked state. (III) A stable mode-locked state can be reached after the pump modulation is turned off. (b-d) DFT measurements of the transitions of states. Note that the time in (b) is scaled by the modulation cycle, instead of the roundtrip time of the cavity, to better reveal self-Q-switching. Besides, we attribute the spikes at the end of the modulation to our pump laser overshooting as the modulation is turned off.}
\label{fig:cartoon}
\end{figure}

To understand the pulse formation, we perform both the DFT and pulse-train measurements for the starting process. Figure \ref{fig:cartoon}(a) illustrates how the state of the laser evolves within the starting procedure. Instead of directly evolving into a modulated mode-locked state, self-Q-switching is observed at the beginning of pump modulation [Fig.~\ref{fig:cartoon}(b)]. After a duration that can span over \SI{200}{\ms}, it spontaneously evolves into a broadband modulated mode-locked state which exhibits a periodically pump-modulated spectrum [Fig.~\ref{fig:cartoon}(c)]. The gain varies on the \SIadj{100}{\us} time scale with \SI{4}{\W} pump power \cite{Lindberg2016}, which is much slower than the pump modulation; therefore, the pulse train can be maintained when the pump is at the minimum. Rather than irregular spectral broadening during the transition of self-Q-switching into a modulated mode-locked state, we see a smooth transition of the spectrum into the stable CW-pumped one [near 1200 roundtrips in Fig.~\ref{fig:cartoon}(d)]. Besides, the laser can reach a stable mode-locked state from a modulated mode-locked state with over \SI{99}{\%} success rate.

\subsection{Path to high-energy pulses}

\subsubsection{Strategies of choosing the path}

\begin{figure}[h!]
\centering\includegraphics[width=3.32in]{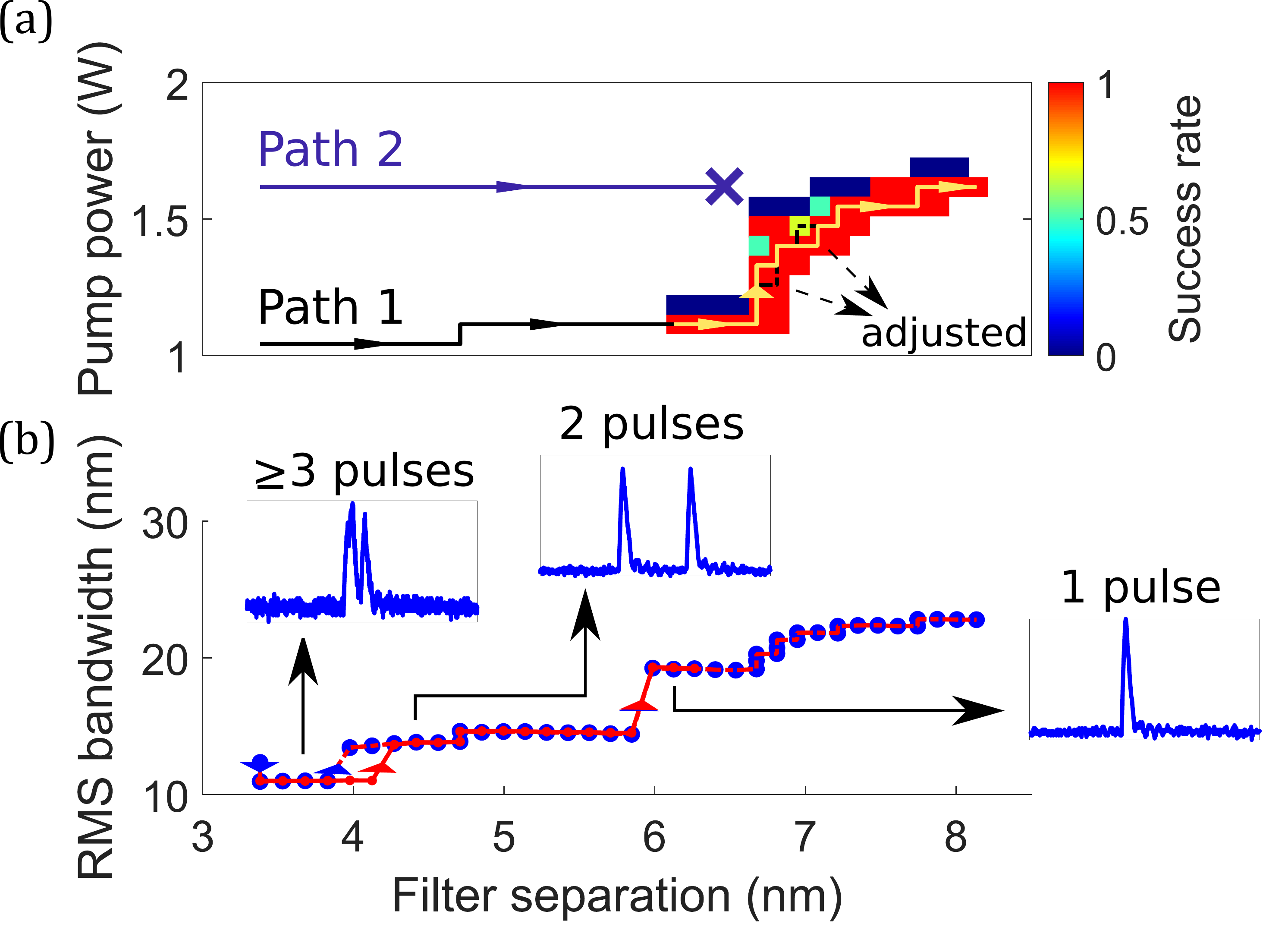}
\caption{(a) The path to a high-energy state after successful starting and (b) the spectral evolution along path 1. Two paths are shown, where path 1 (black and orange lines) can reach the desired state whereas path 2 (blue) fails. Note that we found out that the orange path needed small adjustment after running the entire starting process for more than 300 times (two brown dashed lines). Red and blue lines in (b) are different trials of moving along the path. Three pulse trains, plotted in one roundtrip time, are shown to indicate the number of pulses in the cavity.}
\label{fig:path}
\end{figure}

The initial mode-locked state generally has multiple pulses in the cavity. To obtain a high-energy single-pulse state, the laser is adjusted on a path of increasing pump power and filter separation after successful starting [path 1 in Fig.~\ref{fig:path}(a)]. This path includes two major steps with specific goals: (1) to obtain a single-pulsing state (black line) and (2) to obtain a high-energy state (orange line).

To ensure single-pulse operation, the filter separation is increased in the first step. This forces the laser to operate with a reduced number of pulses in the cavity such that each pulse experiences stronger spectral broadening. During this step, we observe stepwise increases in the spectral bandwidth [Fig.~\ref{fig:path}(b)]. The steps correspond to the transitions into states of fewer pulses. The transition from a stable multi-pulsing mode-locked state to a stable single-pulsing state is successful in \SI{64}{\%} of trials, with failure most often happening where reduction of the number of pulses occurs.

\begin{figure}[h!]
\centering\includegraphics[width=3.32in]{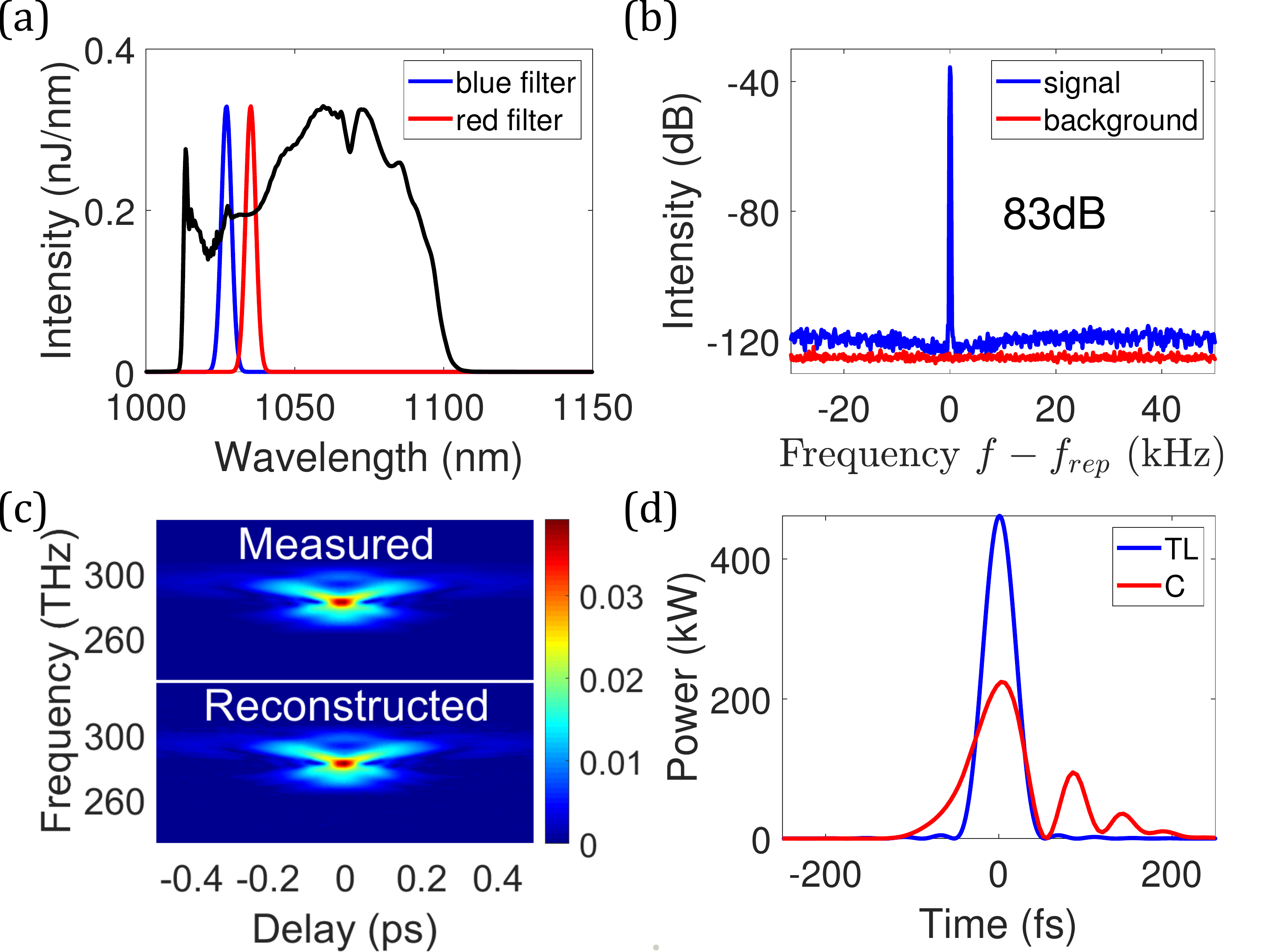}
\caption{(a) The spectra of its delivered pulse and the corresponding filters. (b) Its RF spectrum; $f_{\text{rep}}=\SI{16.1}{\MHz}$: repetition rate of the laser (VBS: \SI{5}{\Hz}, RBS: \SI{10}{\Hz}). (c) The measured and reconstructed FROG trace of the compressed pulse. (d) The temporal profile of the retrieved pulse from the FROG and its transform-limited counterpart. TL: transform-limited pulse; C: compressed pulse.}
\label{fig:result_without_SBS}
\end{figure}

After a single-pulsing state is established, the path finding algorithm is applied to search for the path to the desired high-energy state \cite{Hart1968,Delling2009}. The filter separation is increased from \SI{3.4}{\nm} to \SI{8.1}{\nm} and the pump power from \SI{1.2}{\W} to \SI{1.8}{\W}, beyond which mode-locking is not stable and lost. The final single-pulse state reaches \SIadj{90}{\nm} bandwidth [Fig.~\ref{fig:result_without_SBS}(a)] and \SIadj{21}{\nano\joule} pulse energy. The pulse can be compressed to \SIadj{65}{\fs} FWHM duration with a grating pair [Fig.~\ref{fig:result_without_SBS}(c,d)]. The mode-locked state shows \SIadj{83}{\dB} contrast in its RF spectrum [Fig.~\ref{fig:result_without_SBS}(b)].

\subsubsection{Discussion}

While there are various paths toward the state of desired parameters, some of them aren’t capable of reaching the desired state. Besides the previously discussed path (path 1), we also studied the one of varying only the filter separation while maintaining the targeted pump power (path 2). This type of path was used for the linear Mamyshev oscillator without FRs in Fig.~\ref{fig:result_with_SBS}. and is able to directly achieve a high-energy single-pulse state. However, with the FRs, the mode-locked state is lost at \SIadj{\sim6.5}{\nm} filter separation when laser is still multi-pulsing. The threshold filter separation for single-pulsing is larger than that for mode-locking. Therefore, the pump power needs to be reduced for single-pulsing operation. Path 1 is then followed to successfully obtain a single-pulsing state.

Overall, the starting sequence reaches a stable mode-locked state on \SI{64}{\%} of trials, which is dominated by the success rate of getting single-pulsing. It takes less than 10 seconds to perform the process, so when it fails, it can be repeated. Although the sequence involves controlling and monitoring several parameters of the laser, it can be easily automated and we have succeeded implementing it in our lab.

\section{Numerical simulations}

We obtain further insight into the operation of the linear Mamyshev oscillator by performing numerical simulations. The generalized nonlinear Schrodinger equation is solved in each section of the fibers in conjunction with the rate equations of the Yb-doped fiber \cite{Lindberg2016,Chen2012}. Saturation of the gain by both co- and counter-propagating pulses is considered because Yb ions respond slowly compared to the repetition rate of the laser.

\begin{figure}[h!]
\centering\includegraphics[width=3.32in]{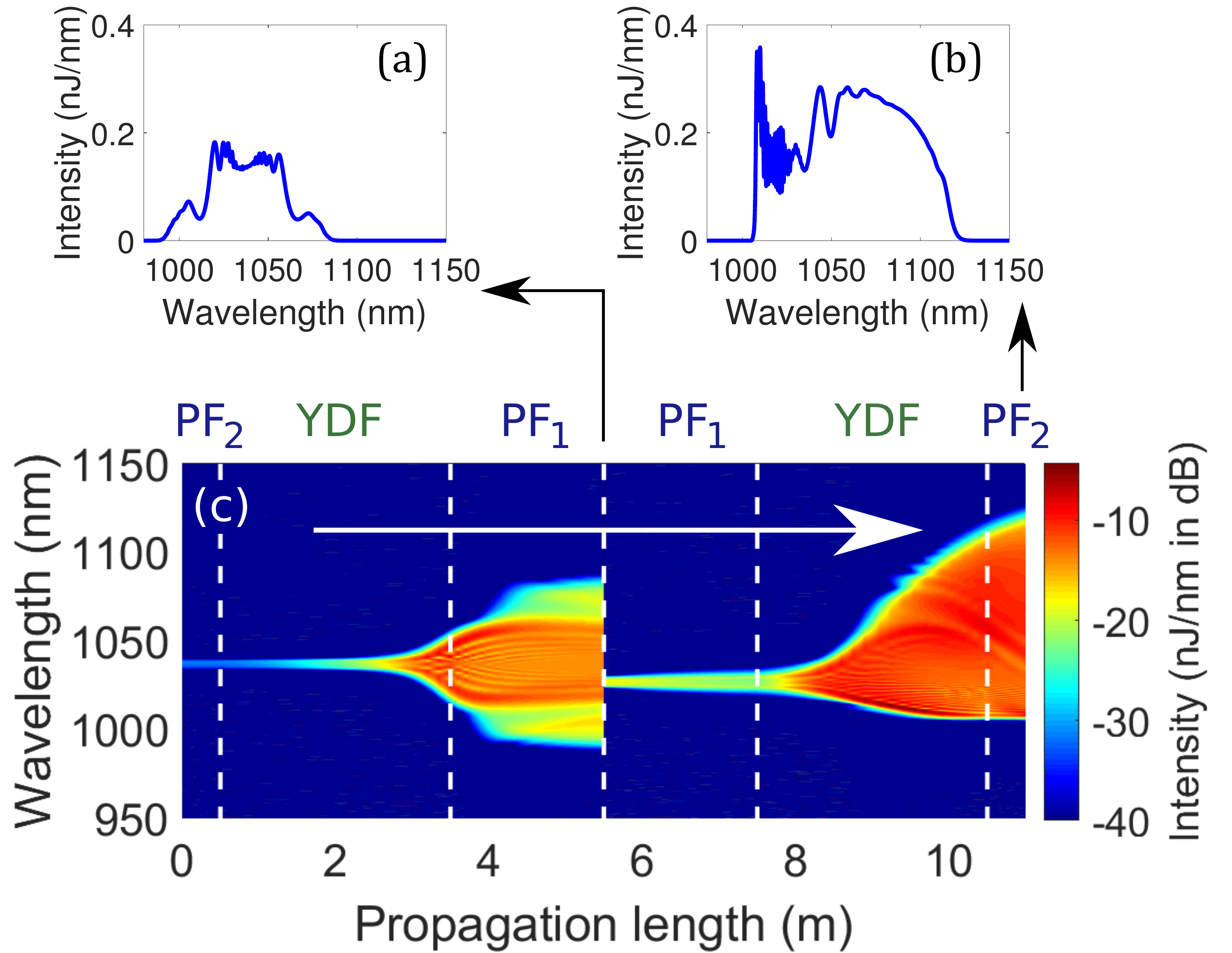}
\caption{The spectra before two filters are plotted out separately (a, b) while the entire spectral evolution within the cavity is also shown (c). Note that the spectral evolution is plotted with decibel (dB).}
\label{fig:simulation}
\end{figure}

Figure \ref{fig:simulation} shows the results with \SIadj{21}{\nano\joule} output pulse energy. The output spectrum [Fig.~\ref{fig:simulation}(b)] is consistent with the experimental results in Fig.~\ref{fig:result_without_SBS}(a). Figure \ref{fig:simulation}(c) shows that the pulse evolves with symmetric broadening by self-phase modulation (SPM) when propagating away from the output coupler [Fig.~\ref{fig:simulation}(a)]; on the other hand, it enters a gain-managed nonlinear (GMN) regime when propagating toward the output [Fig.~\ref{fig:simulation}(b)] \cite{Sidorenko2019}. The spectrum is broadened asymmetrically by following the gain spectrum during the evolution. GMN pulses can accumulate large nonlinear phase and spectral broadening, which allows a short transform-limited pulse duration and maintain good pulse quality. We also realize that, in a linear Mamyshev oscillator, the gain saturation by pulses from both directions results in the SPM regime; otherwise, the two pulses would exhibit similar GMN behavior.

With optimization of fiber lengths and pump power, the simulation predicts stable pulse energy up to 50 nJ and duration of \SI{30}{\fs} after compression. This performance is approached experimentally in our linear Mamyshev oscillator without FRs (Fig.~\ref{fig:result_with_SBS}). We expect to obtain pulses of the similar level in the one with FRs. Because the parameter space of a single-pulsing regime becomes narrow as the pulse energy increases, it may require dense search of the path on the map of pump power and filter separation.

\section{Conclusion}

In conclusion, we have shown that a linear Mamyshev oscillator can deliver \SIadj{40}{\nano\joule} and \SIadj{35}{\fs} pulses. However, it is susceptible to damage of the gain fiber; hence, a new design with two FRs is applied to suppress SBS. The oscillator can be started reliably by modulating the pump power to obtain a mode-locked state, which is then optimized by adjusting the filters and pump power. Full electronic control of this starting procedure is achieved. The laser generates \SIadj{\sim20}{\nano\joule} and \SIadj{65}{\fs} pulses. It offers excellent performance, is environmentally stable, and requires fewer components than a ring laser. We believe that it can be an attractive source for various applications.


\bibliography{references}






\end{document}